\documentstyle[emulateapj,psfig,natbib,amsfonts,amsmath]{article}
\def\ra#1#2#3{#1$^{\rm h}$#2$^{\rm m}$#3$^{\rm s}$}
\def\dec#1#2#3{$#1^\circ#2'#3''$}

\def\grb{GRB\,050904}
\def\swift{{\it Swift}}
\def\spitzer{{\it Spitzer}}

\def\ociw{1}
\def\prince{2}
\def\hubble{3}
\def\ssc{4}
\def\uh{5}
\def\anu{6}
\def\psu{7}
\def\srl{8}
\def\cit{9}
\def\wisc{10}

\begin{document}

\title{{\it HST} and {\it Spitzer} Observations of the Host 
Galaxy of GRB\,050904: A Metal-Enriched, Dusty Starburst at $z=6.295$}

\author{
E.~Berger\altaffilmark{\ociw,}\altaffilmark{\prince,}\altaffilmark{\hubble},
R.~Chary\altaffilmark{\ssc}
L.~L.~Cowie\altaffilmark{\uh}
P.~A.~Price\altaffilmark{\uh},
B.~P.~Schmidt\altaffilmark{\anu},
D.~B.~Fox\altaffilmark{\psu},
S.~B.~Cenko\altaffilmark{\srl},
S.~G.~Djorgovski\altaffilmark{\cit},
A.~M.~Soderberg\altaffilmark{\cit},
S.~R.~Kulkarni\altaffilmark{\cit},
P.~J.~McCarthy\altaffilmark{\ociw},
M.~D.~Gladders\altaffilmark{\ociw,}\altaffilmark{\hubble},
B.~A.~Peterson\altaffilmark{\anu},
and A.~J.~Barger\altaffilmark{\wisc}
}

\altaffiltext{\ociw}{Observatories of the Carnegie Institution
of Washington, 813 Santa Barbara Street, Pasadena, CA 91101}
 
\altaffiltext{\prince}{Princeton University Observatory,
Peyton Hall, Ivy Lane, Princeton, NJ 08544}
 
\altaffiltext{\hubble}{Hubble Fellow}

\altaffiltext{\ssc}{Spitzer Science Center, California Institute 
of Technology, Mail Stop 220-6, Pasadena, CA 91125}

\altaffiltext{\uh}{Institute for Astronomy, University of Hawaii, 
2680 Woodlawn Drive, Honolulu, HI 96822}

\altaffiltext{\srl}{Space Radiation Laboratory, MS 220-47, California
Institute of Technology, Pasadena, CA 91125}
 
\altaffiltext{\cit}{Division of Physics, Mathematics and Astronomy,
105-24, California Institute of Technology, Pasadena, CA 91125}
 
\altaffiltext{\psu}{Department of Astronomy and Astrophysics,
Pennsylvania State University, 525 Davey Laboratory, University
Park, PA 16802}

\altaffiltext{\anu}{RSAA, ANU, Mt.~Stromlo Observatory, via 
Cotter Rd, Weston Creek, ACT 2611, Australia}

\altaffiltext{\wisc}{Department of Astronomy, University of 
Wisconsin-Madison, 475 North Charter Street, Madison, WI 53706}

\begin{abstract} 
We present deep {\it Hubble Space Telescope} and {\it Spitzer Space
Telescope} observations of the host galaxy of \grb\ at $z=6.295$.  The
host is detected in the $H$-band and marginally at $3.6$ $\mu$m.  From
these detections, and limits in the $z'$-band and 4.5 $\mu$m, we infer
an extinction-corrected absolute magnitude, $M_{\rm UV}\approx -20.7$
mag, or $\sim L$*, a substantial star formation rate of $\sim 15$
M$_\odot$ yr$^{-1}$, and a stellar mass of ${\rm few}\times 10^9$
M$_\odot$.  A comparison to the published sample of
spectroscopically-confirmed galaxies at $z>5.5$ reveals that the host
of \grb\ would evade detection and/or confirmation in any of the
current surveys due to the lack of detectable Ly$\alpha$ emission,
which is likely the result of dust extinction ($A_{1200}\sim 1.5$
mag).  This suggests that not all luminous starburst galaxies at
$z\sim 6$ are currently being accounted for.  Most importantly, using
the metallicity of $Z\approx 0.05$ Z$_\odot$ inferred from the
afterglow absorption spectrum, our observations indicate for the first
time that the observed evolution in the mass- and luminosity-metallicity 
relations from $z=0$ to $z\sim 2$ continues on to $z>6$.  The ease of 
measuring redshifts and metallicities from the afterglow emission 
suggests that in tandem with the next generation ground- and space-based 
telescopes, a GRB mission with dedicated near-IR follow-up can provide 
unique information on the evolution of stars and galaxies through the 
epoch of re-ionization.
\end{abstract}
 

\section{Introduction}
\label{sec:intro}

The questions of how and when the universe was re-ionized and the
history of galaxy formation and metal enrichment appear to be
intimately linked.  Observations of $z>6$ quasars and the cosmic
microwave background indicate that re-ionization occurred at
$z\sim 7-13$ \citep{bfw+01,sbd+06}, but most likely not by quasars
alone \citep{fns+02}.  Instead, star-forming galaxies and/or massive
population III stars may have played a dominant role in this process
(e.g., \citealt{yw04}).  To assess this possibility it is essential to
trace the properties and evolution of galaxies and star formation at
$z\gtrsim 6$.  In recent years, large surveys using narrow-band
Ly$\alpha$ and Lyman drop-out selection have uncovered $\sim 100$
candidate $z\gtrsim 5.5$ galaxies (e.g., \citealt{bit+04,dsg+04}), of
which about a half have been confirmed spectroscopically (e.g.,
\citealt{hcm+02,tan+05}).  These surveys provide initial estimates of
the star formation rate density and luminosity function at these
redshifts (e.g., \citealt{bi06}).

Unfortunately, one of the most crucial measurements in the study of
galaxy evolution, the metallicity, is beyond the reach of current
studies, since at $z\gtrsim 5$ the relevant emission
lines\footnotemark\footnotetext{These include [\ion{O}{2}]$\lambda
3727$, [\ion{O}{3}]$\lambda\lambda 4959,5007$, [\ion{N}{2}]$\lambda
6584$, and the hydrogen Balmer lines.} are very weak and are
redshifted to the mid-IR.  Moreover, since spectroscopic confirmation
relies on Ly$\alpha$ emission, which is easily obscured by dust, the
current samples may be intrinsically biased with respect to dust and
metallicity.  The alternative approach of studying damped Ly$\alpha$
absorbers (DLAs) detected against background quasars also appears to 
be limited to $z\lesssim 5$ \citep{pgw+03}, and is moreover biased in
favor of extended halo gas, which at lower redshifts significantly
underestimates the disk metallicities.  As a result, the apparent
evolution in the mass- and luminosity-metallicity relations ($M$-$Z$
and $L$-$Z$) from $z=0$ to $z\sim 2$ (e.g.,
\citealt{kk04,sep+04,sgl+05,esp+06}) cannot be traced to $z>5$, where 
such information should shed light on the initial stages of mass
build-up and metal enrichment.

For many years it has been speculated that GRBs should exist at $z>6$
and can therefore provide an alternative way to study re-ionization
and to select star-forming galaxies.  A truly unique and exciting
aspect of GRBs is that spectroscopy of their bright afterglows can
easily provide a redshift measurement from UV metal absorption
features and/or Ly$\alpha$ absorption, bypassing the reliance on faint
Ly$\alpha$ emission lines.  More importantly, GRB absorption
spectroscopy also provides a measure of the metallicity and kinematics
of the interstellar medium {\it at the location where active star
formation is taking place}, and may potentially provide direct
information on the nature of the massive progenitor itself.  This
powerful probe of the metallicity in star forming galaxies, and its
redshift evolution, is now being routinely used for a rapidly-growing
sample at $z\sim 2-4$ (e.g., \citealt{bpc+05,cpb+05,sve+05}).

The hope of extending this approach to $z>6$ was finally realized with
the discovery of \grb\ at $z=6.295$, and spectroscopic observations of
its afterglow.  Here we present {\it Hubble Space Telescope} and {\it
Spitzer Space Telescope} follow-up observations of the host galaxy of
\grb\ and discuss its properties in the context of the
spectroscopically-confirmed $z>5.5$ galaxies discovered to date.  We
find that the host is a $\sim L$*, low mass, and modestly dusty
starburst galaxy with a high specific star formation rate.  Combining
these results with the absorption-line metallicity, we place the first
point on the $M$-$Z$ and $L$-$Z$ diagrams of $z>5$ galaxies, and find
that the evolution in these relations continues beyond $z\sim 2$.

\section{Observations}
\label{sec:obs}

\grb\ was detected by the \swift\ satellite on 2005, Sep.~4.078 UT
\citep{cmc+06}.  The burst redshift was photometrically estimated to 
be $z\approx 6.2-6.5$ \citep{pcm+05,tac+05,hnr+06}, and was later 
confirmed spectroscopically to be $z=6.295$ \citep{kka+06}, making 
it the highest redshift GRB observed to date.  The afterglow 
absorption spectrum also revealed a damped Ly$\alpha$ absorber with 
${\rm log}\,N({\rm HI})\approx 21.6$, a metallicity of $Z\approx 
0.05$ Z$_\odot$, and appreciable dust depletion 
\citep{tkk+05,kka+06}.

\subsection{{\it Hubble Space Telescope}}

We observed the position of \grb\ with HST as part of a program to
study the host galaxies of $z>6$ GRBs (GO\,10616).  The observations
were performed with the Advanced Camera for Surveys (ACS) on 2005,
Sep.~26.87 UT, and with the Near Infrared Camera and Multi-Object
Spectrometer (NICMOS) on Sep.~27.28 UT.  A total of 4216 and 10240 s
were obtained in the WFC/F850LP and F160W filters, respectively.

We processed the data using the {\tt multidrizzle} routine
\citep{fh02} in the {\tt stsdas} package of IRAF.  The ACS images 
were drizzled using {\tt pixfrac}=0.8 and {\tt pixscale}=1.0,
resulting in a final pixel scale of $0.05\arcsec$ pixel$^{-1}$.  The
NICMOS images were first re-processed with an improved dark frame
created from the HUDF using the IRAF task {\tt calnica} in the {\tt
nicmos} package.  The resulting images were then drizzled using {\tt
pixfrac}=0.7 and {\tt pixscale}=0.5, leading to a final pixel scale of
$0.1\arcsec$ pixel$^{-1}$.

\begin{figure}
\centerline{\psfig{file=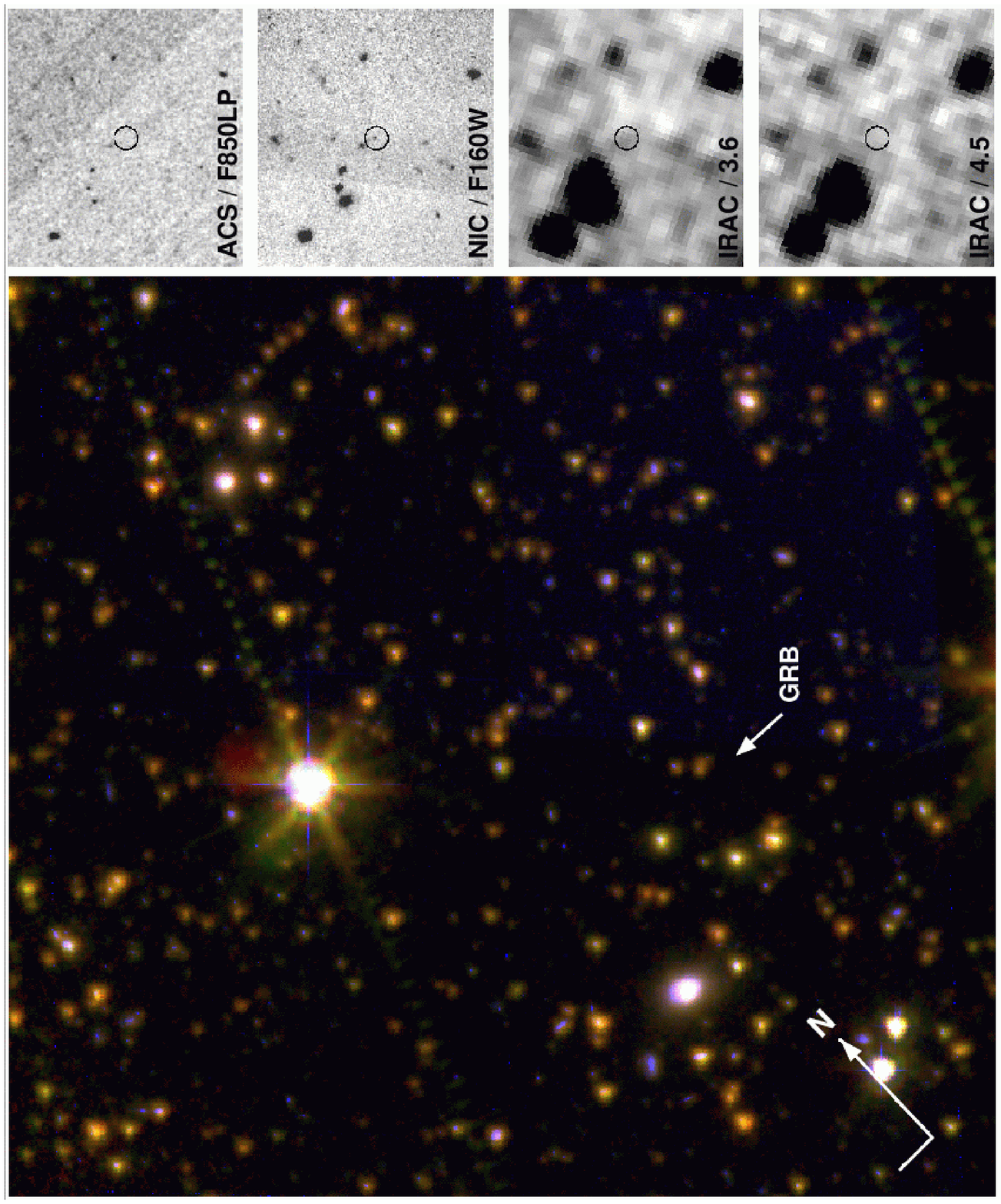,width=6in}}
\caption{A color composite HST+\spitzer\ image of the field of 
GRB\,050904.  The panels on the right provide a zoom-in on the
position of the host galaxy in each of the four available bandpasses.
The host is clearly detected in the NICMOS/F160W image (with an
afterglow contamination of about $50\%$; see \S\ref{sec:obs}), and is
marginally detected at 3.6 $\mu$m.
\label{fig:field}}
\end{figure}

Astrometry was performed relative to a $K$-band image of the afterglow
taken with the Infrared Telescope Facility \citep{hnr+06}.  A total of
five objects in common to the IRTF and NICMOS images were used
resulting in a $1\sigma$ astrometric uncertainty of about
$0.05\arcsec$.  In the NICMOS image we identify a single object
coincident with the afterglow position at (J2000)
$\alpha$=\ra{00}{54}{50.846}, $\delta$=\dec{+14}{05}{09.92}
($0.08\arcsec$ south-east of the afterglow position); see
Figure~\ref{fig:field}.  No corresponding object is detected in the
ACS image.

Photometry of the object was performed using the zero-points of
\citet{sjb+05}, resulting in $m_{\rm AB}({\rm F850LP})>27.0$ mag
($3\sigma$) and $m_{\rm AB}({\rm F160W})=26.1\pm 0.2$ mag ($0.13\pm
0.025$ $\mu$Jy).  An extrapolation of the afterglow flux in the
$H$-band to the time of our observation using the measured decay index
of $\alpha=-2.4$ ($F_\nu\propto\nu^\alpha$; \citealt{tac+05,hnr+06}),
indicates an expected brightness $m_{\rm AB}({\rm F160W})=26.8\pm
0.15$ mag, suggesting that about half of the detected flux is due to
residual afterglow emission.

\begin{figure}
\centerline{\psfig{file=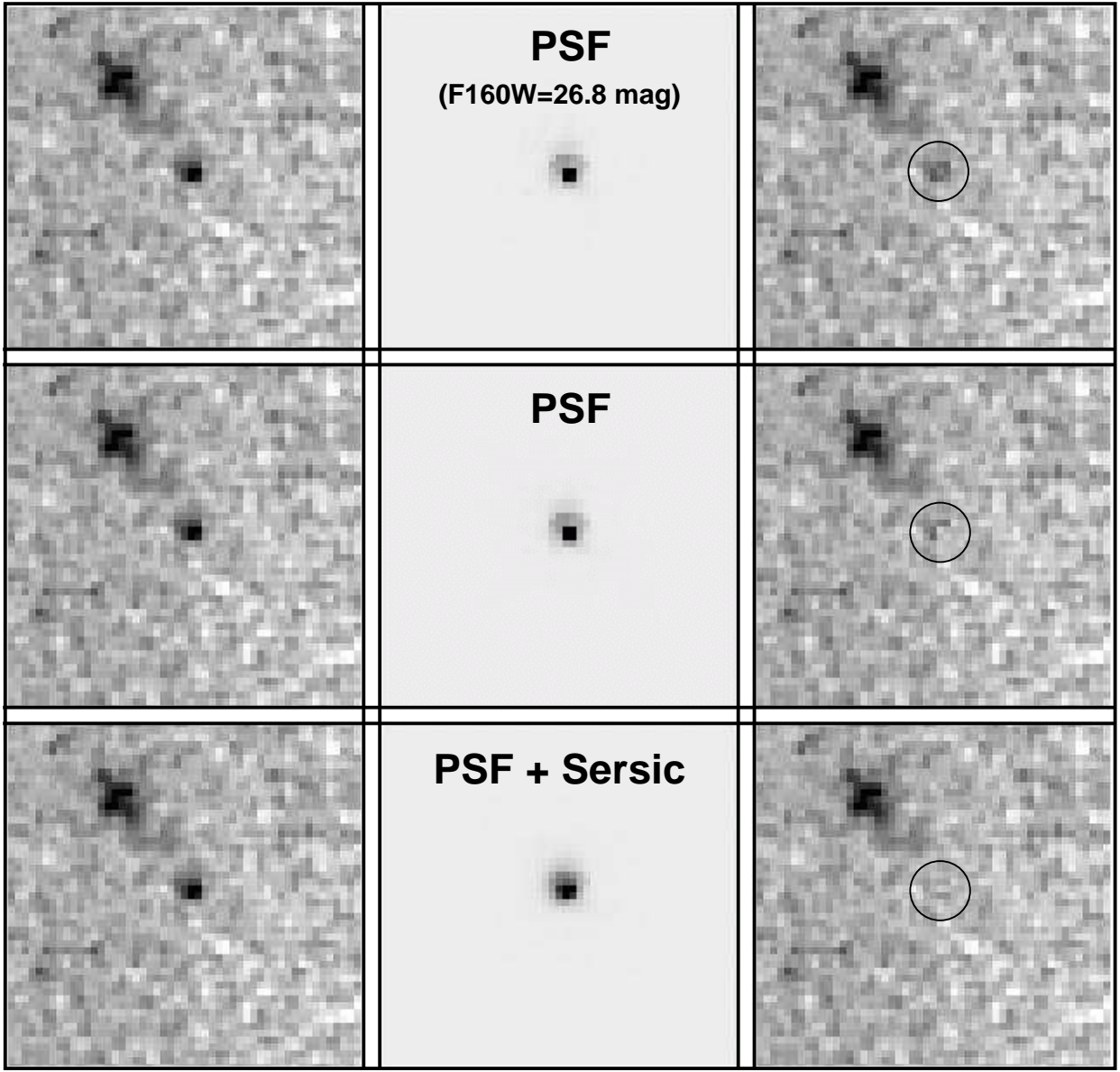,width=6.5in}}
\caption{Model-fitting of the source coincident with \grb\ using 
three models in GALFIT: (Top) point source with $m_{\rm AB}({\rm
F160W})=26.7$ mag as expected from the measured afterglow decay rate;
(middle) point source with the brightness as a free parameter; and
(bottom) a point source with $m_{\rm AB}({\rm F160W})= 26.8$ mag and
an exponential disk with the brightness and scale length as free
parameters.  The middle column shows the resulting model source in
each case, while the right column is the residual image.  Clearly, a
point source alone does not provide an adequate fit, particularly at
the expected flux level.  Instead an extended source which contributes
about half the total flux is required.
\label{fig:galfit}}
\end{figure}

We investigated this in more detail by modeling the surface brightness
of the object using the GALFIT software \citep{phi+02} with a point 
spread function generated with the Tiny Tim
package\footnotemark\footnotetext{\sf
http://www.stsci.edu/software/tinytim/tinytim.html}.  We use three
models: (i) a point-source with $m_{\rm AB}({\rm F160W})=26.8$ mag
appropriate for the predicted afterglow brightness, (ii) a
point-source with the brightness as a free parameter, and (iii) a
point-source with $m_{\rm AB}({\rm F160W})=26.8$ mag and an
exponential disk with the brightness and scale length as free
parameters.  The results of the three fits are shown in
Figure~\ref{fig:galfit}.  We find that models (i) and (ii), for 
which $m_{\rm AB}({\rm F160W})=26.3\pm 0.1$ mag, leave significant 
residuals, indicating the presence of an extended source underlying 
the GRB position.  Model (iii), on the other hand, fully accounts for 
the observed surface brightness profile, resulting in a host galaxy
brightness of $m_{\rm AB}({\rm F160W})=26.7\pm 0.2$ mag and a scale
length of $0.6\pm 0.3\arcsec$.  We therefore conclude that the host
galaxy contributes about $50\%$ of the flux at the time of our
observations.

\subsection{{\it Spitzer Space Telescope}}

As part of the same program (GO\,20000) we also observed the field of
\grb\ with the Infrared Array Camera (IRAC; \citealt{fha+04}) on
\spitzer\ in all four channels (3.6, 4.5, 5.8 and 8.0 $\mu$m) on 2005,
Dec.~25 UT.  The field lies in a region with ``medium''-level zodiacal
background and cirrus of 34 MJy/sr at 24 $\mu$m and 9.6 MJy/sr at 100
$\mu$m on the date of the observations.  We used 100 s integrations
with 72 medium scale dithers from the random cycling pattern for total
on-source integration times of 7200 s at each passband.  These 
nominal $3\sigma$ point source sensitivity limits are 0.26, 0.49, 3.3
and 4.2 $\mu$Jy, respectively.
 
Starting with the S13.2.0 pipeline processed basic calibrated data
(BCD) sets we corrected the individual frames for muxbleed and column
pull down using software developed for the Great Observatories Origins
Deep Survey.  Due to the presence of bright stars in the field, many
of the frames at 3.6 and 4.5 $\mu$m also showed evidence for
``muxstriping''.  This was removed using an additive correction on a
column by column basis (J.~Surace, priv.~comm.).  The processed BCD
frames were then mosaicked together using the MOPEX routine
\citep{mm05} and drizzled onto a $0.6\arcsec$ grid.  Astrometry was
performed relative to the HST/ACS image using 70 common objects,
resulting in an rms uncertainty of $0.07\arcsec$ (3.6 $\mu$m) and
$0.09\arcsec$ (4.5 $\mu$m).

Photometry at the position of the host galaxy was performed in fixed
circular apertures of $1.2\arcsec$ radius with appropriate beam size
corrections applied as stated in the \spitzer\ Observer's Manual.  The
presence of brighter sources within $\sim 7\arcsec$ of the host
position required that we fit for the wings of the point spread
function from those sources.  We find a marginal detection of the host
with $0.17\pm 0.09$ $\mu$Jy at 3.6 $\mu$m, and $3\sigma$ upper limits
of 0.4 $\mu$Jy at 4.5 $\mu$m, 2.7 $\mu$Jy at 5.8 $\mu$m, and 2.5
$\mu$Jy at 8.0 $\mu$m.  With the observed afterglow spectral index,
$\beta_\nu=-1.25$ ($F_\nu\propto\nu^{\beta_\nu}$;
\citealt{tac+05,hnr+06}), we find that the expected 3.6 $\mu$m afterglow
flux at the time of our observation is negligible, $\lesssim 0.005$
$\mu$Jy.

\section{Host Galaxy Properties} 
\label{sec:prop}

\begin{figure}
\centerline{\psfig{file=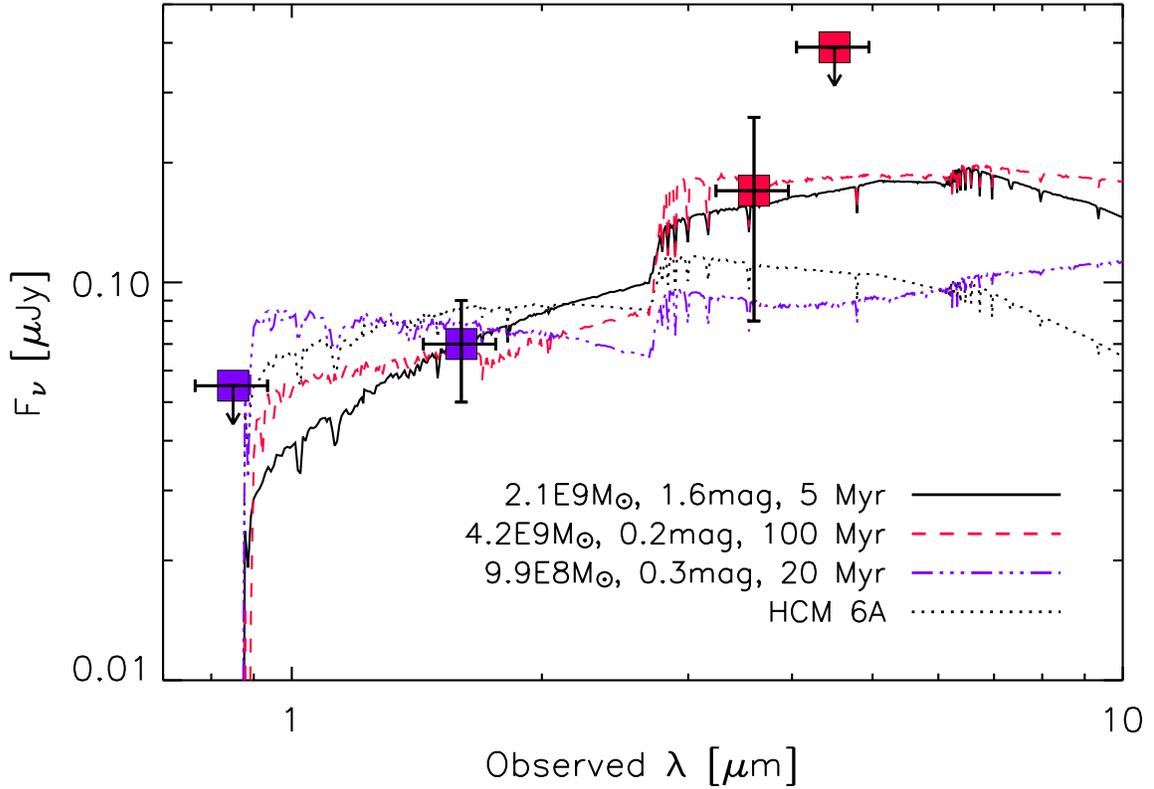,width=6.5in}}
\caption{Spectral energy distribution of the host galaxy of \grb\ 
from HST (blue) and \spitzer\ (red) data.  Three representative SEDs
are shown (see \S\ref{sec:prop} for details) with model parameters
given in the figure.  The models with $A_V\sim 0.2-0.3$ mag are based
on the extinction inferred from the afterglow emission.  For
comparison, the dotted line represents the best-fit model to the SED
of the $z=6.56$ galaxy HCM\,6A (redshifted to $z=6.295$) with an age
of 5 Myr, $A_V=1.0$ mag, and $M=8.4\times 10^8$ M$_\odot$
\citep{cse05}.
\label{fig:sed}}
\end{figure}

Using the observed fluxes and upper limits (Figure~\ref{fig:sed}) we
now investigate the physical properties of the host
galaxy\footnotemark\footnotetext{We use the standard cosmological
parameters: ${\rm H}_0=70$ km s$^{-1}$ Mpc$^{-1}$, $\Omega_m=0.27$,
and $\Omega_\Lambda=0.73$, which lead to $d_L=1.95\times 10^{29}$ cm
and $1\arcsec=5.76$ kpc at $z=6.295$.}.  We begin by estimating the
host extinction using (i) the difference between the observed and
expected\footnotemark\footnotetext{The value of $-0.55$ is inferred
from the optical time decay rate and the typical assumption of
$\nu_m<\nu_{\rm opt}<\nu_c$, where $\nu_m$ and $\nu_c$ are the
synchrotron peak and cooling frequencies, respectively.} afterglow
spectral indices, $\beta_\nu=-1.25$ and $-0.55$, respectively
\citep{tac+05,hnr+06}, and (ii) the host UV slope (the
$\beta_\lambda$-$A_{1600}$ relation; \citealt{mhc99}).  The former
approach indicates that for a \citet{cal97} extinction curve
$A_V\approx 0.3$ mag, or at the observed F160W and F850LP bandpasses,
$A_{2200}\approx 0.7$ mag and $A_{1400}\approx 1.2$ mag.  With the
latter approach we find $\beta_\lambda\gtrsim -1.5$ and hence
$A_{1600}=4.43+1.99\beta_\lambda\gtrsim 1.4$ mag, in rough agreement
with the afterglow-based results; here $\beta_\lambda$ is defined such
that $F_\lambda\propto\lambda^{\beta_\lambda}$.  We note that the
extinction estimates agree with the significant dust depletion
inferred from the afterglow absorption spectrum \citep{kka+06}.

At the redshift of the host the observed 3.6 $\mu$m band roughly
traces the rest-frame optical $B$-band, leading to an
extinction-corrected absolute magnitude, $M_{\rm AB}(B)\approx -20.7$
mag.  The rest-frame UV magnitude, traced by the observed F160W band,
is $M_{\rm AB}(2200)\approx -20.8$ mag, or $M_{\rm AB}(1400)
\gtrsim -20.7$ mag if we use the F850LP limit.  These values
correspond to a luminosity, $L\approx 1.5L$* compared to the
luminosity function of $z\sim 6$ candidates in the HUDF (based on
photometric redshifts alone; \citealt{bi06}), or about $0.7L$* when
compared to the luminosity function of $z\sim 3-4$ Lyman break
galaxies (LBGs; \citealt{sag+99}).  Thus, the host of \grb\ is roughly
an $L$* galaxy.

The host star formation rate (SFR) can be estimated from the measured
UV luminosity and the conversion relation of \citet{ken98}.  Based on
the F160W flux, we find $L_\nu (2200)=(4.6\pm 1.3)\times 10^{28}$ erg
s$^{-1}$ Hz$^{-1}$, or ${\rm SFR}=6.5\pm 1.8$ M$_\odot$ yr$^{-1}$.
The extinction-corrected rate is about $15$ M$_\odot$ yr$^{-1}$ using
the value of $A_{2200}$ inferred above.  These values can be
contrasted with the limit of $\lesssim 0.8$ M$_\odot$ yr$^{-1}$
inferred from the lack of detectable Ly$\alpha$ emission in the
absorption spectrum of \grb\ \citep{tkk+05}.  The discrepancy can be
easily explained in terms of dust absorption of the Ly$\alpha$
photons, providing additional support to the significant extinction
inferred from the afterglow spectral slope, the host UV spectral
slope, and the depletion pattern.

Incorporating the \spitzer\ data, we provide rough constraints on the
stellar mass of the host galaxy.  Given the low signal-to-noise we do
not attempt real model fits, but instead we adopt a $Z=0.2$ Z$_\odot$,
Salpeter IMF template from the \citet{bc03} library along with
representative values of the mass, stellar population age, and
extinction.  An IGM absorption model \citep{mad95} has also been
incorporated in the SED.  The models, shown in Figure~\ref{fig:sed},
indicate that for $A_V\sim 0.3$ mag, as inferred above, the stellar
mass ranges from about $10^9$ to $4\times 10^9$ M$_\odot$ as we vary
the stellar population age from 20 to 100 Myr and span the range of
uncertainty in the $3.6$ $\mu$m flux.  For a larger extinction,
$A_V\sim 1.6$ mag and an age of 5 Myr we find a mass of about $2\times
10^9$ M$_\odot$.  We therefore conclude that the stellar mass is $\sim
{\rm few}\times 10^9$ M$_\odot$, similar for example to the value
derived for the $z=6.56$ galaxy HCM 6A \citep{cse05}, while the
stellar population age is likely $10-100$ Myr.  

Finally, the inferred scale length for the host galaxy based on the
GALFIT model presented in \S\ref{sec:obs} is $r_e=3.4\pm 1.7$ kpc,
consistent with the median value of about 2 kpc for GRB host galaxies
at $\langle z\rangle\sim 1.5$ \citep{wbp05}, as well as with the
median value of about 1 kpc for $i$-band drop-outs in the HUDF
\citep{bib+04}.

To summarize, we find that the host galaxy of \grb\ is an $L$*
starburst galaxy with a mass similar to that of the LMC, with
interstellar gas enriched to about 0.05 Z$_\odot$ \citep{kka+06}, and
with appreciable dust depletion and extinction.

\section{Discussion and Future Directions}
\label{sec:conc}

\begin{figure}
\centerline{\psfig{file=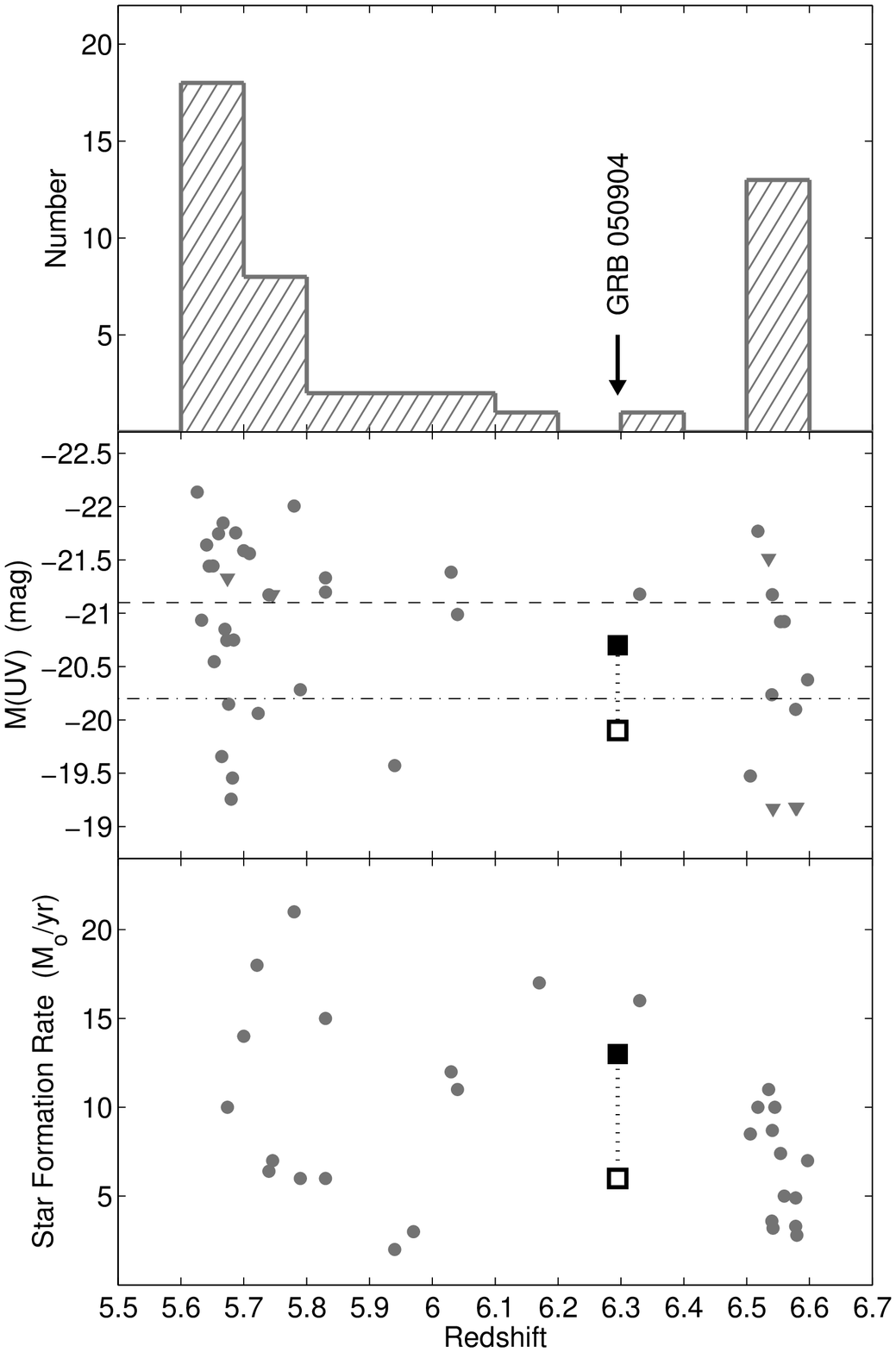,width=5in}}
\caption{The inferred properties of the host of \grb\ compared to 
the published sample of spectroscopically-confirmed galaxies at
$z>5.5$
\citep{hmc99,hcm+02,bse+03,clm+03,ktk+03,rdm+03,dsg+04,hcp+04,kcs+04,nkm+04,rxd+04,sgb+04,cse05,ebs+05,nkm+05,sye+05,sdp+05,tan+05,wjl+05}.  Open and filled black squares 
designate raw and extinction-corrected quantities, respectively, for
the host galaxy.  Both detections (circles) and upper limits
(triangles) are shown for the distributions of redshifts, absolute
rest-frame UV magnitudes, and star formation rates.  The dashed line
in the middle panel designates an $M$* galaxy at $z\sim 3-4$
\citep{sag+99}, while the dash-dotted line is $M$* for $z\sim 6$ 
candidates in the HUDF \citep{bi06}.
\label{fig:gals}}
\end{figure}

\grb\ is by far the highest redshift spectroscopically-confirmed burst
observed to date, and its host is so far the only $z>5$ galaxy for
which an estimate of the metallicity is available.  Given the
relatively small number of spectroscopically-confirmed galaxies at
$z>5.5$, it is instructive to compare the properties of a GRB-selected
galaxy to those selected through narrow-band Ly$\alpha$ imaging or the
Lyman drop-out technique.  In Figure~\ref{fig:gals} we compare some of
the basic properties, which are available for the latter samples,
namely the rest-frame absolute magnitudes and UV/Ly$\alpha$ star
formation rates.  We find that in the published sample, only 14
galaxies are located at higher redshift than the host of \grb.  In
terms of UV luminosity, the host of \grb\ is fainter than about $85\%$
of all the known galaxies, and it has a star formation rate lower than
about $70\%$ of all the galaxies.

On the other hand, we stress that the host of \grb\ would evade
detection or confirmation in the current surveys since its Ly$\alpha$
line flux is very low, $\lesssim 1.7\times 10^{-18}$ erg cm$^{-2}$
s$^{-1}$ \AA$^{-1}$ \citep{kka+06}.  This is most likely due to
absorption by dust as inferred from the afterglow spectral index, the
host UV slope, and dust depletion in the absorption spectrum.  And yet
the host of \grb\ is roughly an $L$* galaxy with an appreciable star
formation rate (as evidenced by the UV luminosity and the occurrence
of a GRB) and a substantial gas reservoir as inferred from the neutral
hydrogen column density of ${\rm log}\,N({\rm HI})\approx 21.6$.
Naturally, a much larger sample is required to accurately assess the
relative contribution of similar galaxies to the star formation budget
at $z\sim 6$, but it is clear that with the ability to measure
redshifts independent of Ly$\alpha$ emission, GRBs may provide the
cleanest handle on obscured star formation at these redshifts.

We now turn to a comparison of the metallicity, mass and luminosity of
the host of \grb\ to those of lower redshift galaxies in an attempt to
provide a first test of evolution in the $M$-$Z$ and $L$-$Z$ relations
from $z\sim 1-2$ to $z>5$.  First, we provide a note of caution that
we are comparing a metallicity derived from absorption lines (in this
case [S/H] since sulfur is a non-refractory element) to the oxygen
abundance derived from emission lines using the $R_{23}$ and $N2$
methods.  In the case of quasar DLAs metallicities a nearly 1 dex
discrepancy has been noted compared to emission line metallicities.
However, unlike quasar sight lines which preferentially probe halo
gas, GRB sight lines probe star forming regions, much like emission
lines diagnostics.  This is supported by observations of
systematically higher metallicities as a function of redshift for
GRB-DLAs compared to QSO-DLAs \citep{bpc+05}. Thus, the comparison to
emission line metallicities is most likely robust, and the only
remaining caveat is that the inferred metallicity potentially
represents the star forming region local to the GRB, and not the
average metallicity of all \ion{H}{2} regions in the galaxy.  In the
absence of additional information, we take the inferred metallicity of
${\rm [S/H]}=-1.3\pm 0.3$ to be representative.

\begin{figure}
\centerline{\psfig{file=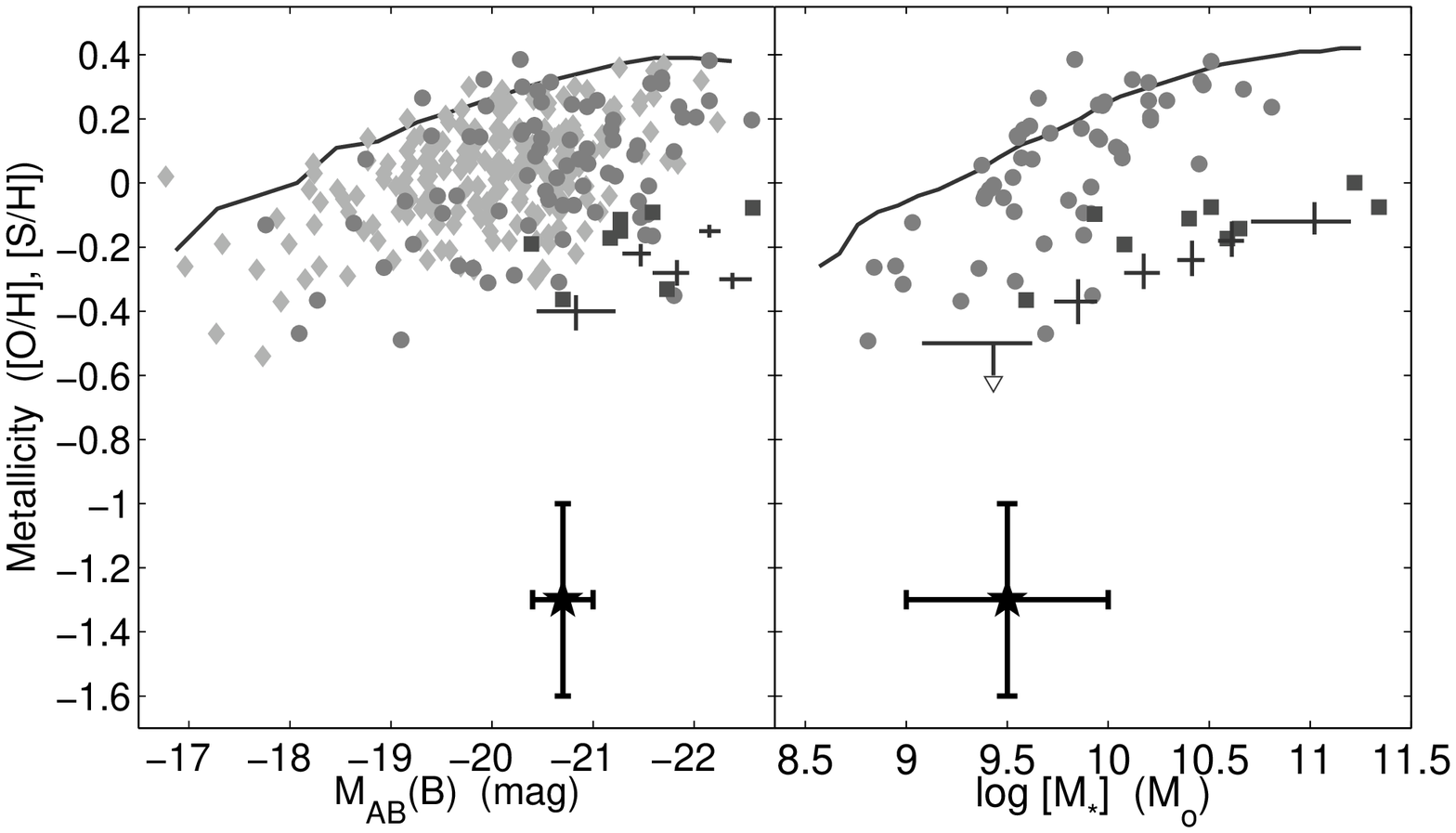,width=6.5in}}
\caption{Metallicity as a function of luminosity (left) and mass 
(right) for the host of \grb\ using the [S/H] value inferred from the
afterglow absorption spectrum \citep{kka+06}.  Also shown are emission
line oxygen abundances for galaxies from GDDS and CFRS at $z\sim
0.4-1.0$ (circles; \citealt{sgl+05}), TKRS at $z\sim 0.3-1.0$
(diamonds; \citealt{kk04}), the DEEP2 survey at $z\sim 1.0-1.5$
(squares; \citealt{scm+05}), and a compilation of 87 LBGs at $z\sim
2.3$ (error bars; \citealt{esp+06}).  The gray lines represent the
relations derived for $z\sim 0.1$ galaxies in the SDSS \citep{thk+04}.
\label{fig:mlz}}
\end{figure}

In Figure~\ref{fig:mlz} we plot the metallicity of the host of \grb\
versus luminosity and stellar mass as inferred in \S\ref{sec:prop}.
For comparison we plot the same data for $z\sim 0.1$ galaxies in the
Sloan Digital Sky Survey \citep{thk+04}, $z\sim 0.3-1.0$ galaxies from
the Gemini Deep Deep Survey, the Canada-France Redshift Survey, and
the Team Keck Redshift Survey \citep{kk04,sgl+05}, $z\sim 1.0-1.5$
galaxies from the DEEP2 survey \citep{scm+05}, and $z\sim 2.3$
UV-selected galaxies \citep{esp+06}.  As noted by the aforementioned
authors, there is clear evolution in both the $M$-$Z$ and $L$-$Z$
relations in the sense that galaxies of a given mass/luminosity have
lower metallicities at progressively higher redshifts.  The
implications of this evolution, and of the $M$-$Z$ and $L$-$Z$
relations themselves, are a matter of current investigation, and here
we simply note that the host of \grb\ indicates that this trend likely
continues to much higher redshifts.

Specifically, for galaxies of a similar brightness to the host of
\grb\ the mean metallicity evolves from $\sim 2$ Z$_\odot$ at $z
\sim 0.1$ to $\sim 1$ Z$_\odot$ at $z\sim 0.7$ and $0.4$ Z$_\odot$ at
$z\sim 2.3$.  A similar trend is observed with mass.  The host of
\grb, with $Z\sim 0.05$ Z$_\odot$ continues this trend.  In fact, the
data for \grb\ are in good agreement with the empirical time evolution
model of the $M$-$Z$ relation derived by \citet{sgl+05}, which
indicates that for ${\rm log}\,M\sim 9.5$ the expected metallicity at
$z=6.3$ is about $0.1$ Z$_\odot$.

The detection of additional GRBs at $z\sim 6$ will allow us to examine
in detail whether the $M$-$Z$ and $L$-$Z$ relations actually exist at
those redshifts, and if they in fact follow the evolutionary trend
observed at lower redshifts.  Moreover, with the ability to probe
galactic-scale outflows in absorption, we can determine whether the
origin of these relations is rooted in higher gas fractions for lower
mass galaxies, or outflows from their shallower potential wells --- a
matter of current debate \citep{mb97,thk+04,esp+06}.  This applies to
the growing sample of GRB absorption spectra at $z\sim 2-4$ as well,
which can both fill in the gap from $z\sim 2$ to $z\sim 6$, and
through near-IR spectroscopy of the host galaxies allow us to compare
emission- and absorption-derived metallicities at $z\sim 2-3$, and
directly assess the existence of any systematic differences.

We conclude by noting that GRB selection of $z\sim 6$ galaxies is
roughly as efficient as other techniques: \grb\ required about 6 hr
for the afterglow identification and spectroscopy
\citep{tac+05,hnr+06,kka+06}, and 6 HST orbits, compared to an average 
of $\sim 10$ hr of large telescope time to locate and confirm a $z\sim
6$ galaxy in other surveys \citep{bit+04,hcp+04,tan+05}.  As
demonstrated in this paper, the real power of GRB selection lies in
the relative ease of redshift determination and the reduced influence
of dust compared to the reliance on faint Ly$\alpha$ emission, and
even more importantly the ability to measure metallicities (and at
high spectral resolution, kinematics).  With this in mind, we suggest
that along with the development of future $20-30$ m ground-based
telescopes and {\it JWST}, a next-generation GRB mission with higher
sensitivity and all-sky coverage, coupled with dedicated near-IR
imaging and spectroscopy follow-up from the ground, may provide a
complementary window into the evolution and metallicity of the first
stars and galaxies.

\acknowledgements 

We thank Lisa Kewley and Alice Shapley for helpful comments on the
manuscript.  EB acknowledges support by NASA through Hubble Fellowship 
grant HST-01171.01 awarded by STSCI, which is operated by AURA, Inc., 
for NASA under contract NAS5-26555.  Additional support was provided by
NASA through grant HST-GO-10616 from STSCI and through a Spitzer award
from JPL/Caltech.

\clearpage

\clearpage

\clearpage

\clearpage

\clearpage


\begin{thebibliography}{}
 
\bibitem[\protect\citeauthoryear{{Becker} et~al.}{{Becker}
  et~al.}{2001}]{bfw+01}
{Becker}, R.~H., et~al. 2001, \aj, 122, 2850
 
\bibitem[\protect\citeauthoryear{{Berger} et~al.}{{Berger}
  et~al.}{2005}]{bpc+05}
{Berger}, E., {Penprase}, B.~E., {Cenko}, S.~B., {Kulkarni}, S.~R., {Fox},
  D.~B., {Steidel}, C.~C.,  \& {Reddy}, N.~A. 2005, ArXiv Astrophysics e-prints
 
\bibitem[\protect\citeauthoryear{{Bouwens} \& {Illingworth}}{{Bouwens} \&
  {Illingworth}}{2006}]{bi06}
{Bouwens}, R.,  \& {Illingworth}, G. 2006, New Astronomy Review, 50, 152
 
\bibitem[\protect\citeauthoryear{{Bouwens} et~al.}{{Bouwens}
  et~al.}{2004a}]{bib+04}
{Bouwens}, R.~J., {Illingworth}, G.~D., {Blakeslee}, J.~P., {Broadhurst},
  T.~J.,  \& {Franx}, M. 2004a, \apjl, 611, L1
 
\bibitem[\protect\citeauthoryear{{Bouwens} et~al.}{{Bouwens}
  et~al.}{2004b}]{bit+04}
{Bouwens}, R.~J., et~al. 2004b, \apjl, 606, L25
 
\bibitem[\protect\citeauthoryear{{Bruzual} \& {Charlot}}{{Bruzual} \&
  {Charlot}}{2003}]{bc03}
{Bruzual}, G.,  \& {Charlot}, S. 2003, \mnras, 344, 1000
 
\bibitem[\protect\citeauthoryear{{Bunker} et~al.}{{Bunker}
  et~al.}{2003}]{bse+03}
{Bunker}, A.~J., {Stanway}, E.~R., {Ellis}, R.~S., {McMahon}, R.~G.,  \&
  {McCarthy}, P.~J. 2003, \mnras, 342, L47
 
\bibitem[\protect\citeauthoryear{{Calzetti}}{{Calzetti}}{1997}]{cal97}
{Calzetti}, D. 1997, \aj, 113, 162
 
\bibitem[\protect\citeauthoryear{{Chary}, {Stern}, \& {Eisenhardt}}{{Chary}
  et~al.}{2005}]{cse05}
{Chary}, R.-R., {Stern}, D.,  \& {Eisenhardt}, P. 2005, \apjl, 635, L5
 
\bibitem[\protect\citeauthoryear{{Chen} et~al.}{{Chen} et~al.}{2005}]{cpb+05}
{Chen}, H.-W., {Prochaska}, J.~X., {Bloom}, J.~S.,  \& {Thompson}, I.~B. 2005,
  \apjl, 634, L25
 
\bibitem[\protect\citeauthoryear{{Cuby} et~al.}{{Cuby} et~al.}{2003}]{clm+03}
{Cuby}, J.-G., {Le F{\`e}vre}, O., {McCracken}, H., {Cuillandre}, J.-C.,
  {Magnier}, E.,  \& {Meneux}, B. 2003, \aap, 405, L19

\bibitem[\protect\citeauthoryear{{Cusumano} et~al.}{{Cusumano}
  et~al.}{2006}]{cmc+06}
{Cusumano}, G., et~al. 2006, \nat, 440, 164
 
\bibitem[\protect\citeauthoryear{{Dickinson} et~al.}{{Dickinson}
  et~al.}{2004}]{dsg+04}
{Dickinson}, M., et~al. 2004, \apjl, 600, L99
 
\bibitem[\protect\citeauthoryear{{Erb} et~al.}{{Erb} et~al.}{2006}]{esp+06}
{Erb}, D.~K., {Shapley}, A.~E., {Pettini}, M., {Steidel}, C.~C., {Reddy},
  N.~A.,  \& {Adelberger}, K.~L. 2006, ArXiv Astrophysics e-prints
 
\bibitem[\protect\citeauthoryear{{Eyles} et~al.}{{Eyles} et~al.}{2005}]{ebs+05}
{Eyles}, L.~P., {Bunker}, A.~J., {Stanway}, E.~R., {Lacy}, M., {Ellis}, R.~S.,
  \& {Doherty}, M. 2005, \mnras, 364, 443
 
\bibitem[\protect\citeauthoryear{{Fan} et~al.}{{Fan} et~al.}{2002}]{fns+02}
{Fan}, X., {Narayanan}, V.~K., {Strauss}, M.~A., {White}, R.~L., {Becker},
  R.~H., {Pentericci}, L.,  \& {Rix}, H.-W. 2002, \aj, 123, 1247
 
\bibitem[\protect\citeauthoryear{{Fazio} et~al.}{{Fazio} et~al.}{2004}]{fha+04}
{Fazio}, G.~G., et~al. 2004, \apjs, 154, 10
 
\bibitem[\protect\citeauthoryear{{Fruchter} \& {Hook}}{{Fruchter} \&
  {Hook}}{2002}]{fh02}
{Fruchter}, A.~S.,  \& {Hook}, R.~N. 2002, \pasp, 114, 144
 
\bibitem[\protect\citeauthoryear{{Haislip} et~al.}{{Haislip}
  et~al.}{2006}]{hnr+06}
{Haislip}, J.~B., et~al. 2006, \nat, 440, 181
 
\bibitem[\protect\citeauthoryear{{Hu} et~al.}{{Hu} et~al.}{2004}]{hcp+04}
{Hu}, E.~M., {Cowie}, L.~L., {Capak}, P., {McMahon}, R.~G., {Hayashino}, T.,
  \& {Komiyama}, Y. 2004, \aj, 127, 563
 
\bibitem[\protect\citeauthoryear{{Hu} et~al.}{{Hu} et~al.}{2002}]{hcm+02}
{Hu}, E.~M., {Cowie}, L.~L., {McMahon}, R.~G., {Capak}, P., {Iwamuro}, F.,
  {Kneib}, J.-P., {Maihara}, T.,  \& {Motohara}, K. 2002, \apjl, 568, L75
 
\bibitem[\protect\citeauthoryear{{Hu}, {McMahon}, \& {Cowie}}{{Hu}
  et~al.}{1999}]{hmc99}
{Hu}, E.~M., {McMahon}, R.~G.,  \& {Cowie}, L.~L. 1999, \apjl, 522, L9
 
\bibitem[\protect\citeauthoryear{{Kawai} et~al.}{{Kawai} et~al.}{2006}]{kka+06}
{Kawai}, N., et~al. 2006, \nat, 440, 184
 
\bibitem[\protect\citeauthoryear{{Kennicutt}}{{Kennicutt}}{1998}]{ken98}
{Kennicutt}, R.~C. 1998, \araa, 36, 189
 
\bibitem[\protect\citeauthoryear{{Kobulnicky} \& {Kewley}}{{Kobulnicky} \&
  {Kewley}}{2004}]{kk04}
{Kobulnicky}, H.~A.,  \& {Kewley}, L.~J. 2004, \apj, 617, 240
 
\bibitem[\protect\citeauthoryear{{Kodaira} et~al.}{{Kodaira}
  et~al.}{2003}]{ktk+03}
{Kodaira}, K., et~al. 2003, \pasj, 55, L17
 
\bibitem[\protect\citeauthoryear{{Kurk} et~al.}{{Kurk} et~al.}{2004}]{kcs+04}
{Kurk}, J.~D., {Cimatti}, A., {di Serego Alighieri}, S., {Vernet}, J., {Daddi},
  E., {Ferrara}, A.,  \& {Ciardi}, B. 2004, \aap, 422, L13
 
\bibitem[\protect\citeauthoryear{{Madau}}{{Madau}}{1995}]{mad95}
{Madau}, P. 1995, \apj, 441, 18
 
\bibitem[\protect\citeauthoryear{{Makovoz} \& {Marleau}}{{Makovoz} \&
  {Marleau}}{2005}]{mm05}
{Makovoz}, D.,  \& {Marleau}, F.~R. 2005, \pasp, 117, 1113
 
\bibitem[\protect\citeauthoryear{{McGaugh} \& {de Blok}}{{McGaugh} \& {de
  Blok}}{1997}]{mb97}
{McGaugh}, S.~S.,  \& {de Blok}, W.~J.~G. 1997, \apj, 481, 689
 
\bibitem[\protect\citeauthoryear{{Meurer}, {Heckman}, \& {Calzetti}}{{Meurer}
  et~al.}{1999}]{mhc99}
{Meurer}, G.~R., {Heckman}, T.~M.,  \& {Calzetti}, D. 1999, \apj, 521, 64
 
\bibitem[\protect\citeauthoryear{{Nagao} et~al.}{{Nagao} et~al.}{2005}]{nkm+05}
{Nagao}, T., et~al. 2005, \apj, 634, 142
 
\bibitem[\protect\citeauthoryear{{Nagao} et~al.}{{Nagao} et~al.}{2004}]{nkm+04}
{Nagao}, T., {Kawabata}, K.~S., {Murayama}, T., {Ohyama}, Y., {Taniguchi}, Y.,
  {Sumiya}, R.,  \& {Sasaki}, S.~S. 2004, \aj, 128, 109
 
\bibitem[\protect\citeauthoryear{{Peng} et~al.}{{Peng} et~al.}{2002}]{phi+02}
{Peng}, C.~Y., {Ho}, L.~C., {Impey}, C.~D.,  \& {Rix}, H. 2002, \aj, 124, 266
 
\bibitem[\protect\citeauthoryear{{Price} et~al.}{{Price} et~al.}{2005}]{pcm+05}
{Price}, P.~A., {Cowie}, L.~L., {Minezaki}, T., {Schmidt}, B.~P., {Songaila},
  A.,  \& {Yoshii}, Y. 2005, ArXiv Astrophysics e-prints
 
\bibitem[\protect\citeauthoryear{{Prochaska} et~al.}{{Prochaska}
  et~al.}{2003}]{pgw+03}
{Prochaska}, J.~X., {Gawiser}, E., {Wolfe}, A.~M., {Castro}, S.,  \&
  {Djorgovski}, S.~G. 2003, \apjl, 595, L9
 
\bibitem[\protect\citeauthoryear{{Rhoads} et~al.}{{Rhoads}
  et~al.}{2003}]{rdm+03}
{Rhoads}, J.~E., et~al. 2003, \aj, 125, 1006
 
\bibitem[\protect\citeauthoryear{{Rhoads} et~al.}{{Rhoads}
  et~al.}{2004}]{rxd+04}
{Rhoads}, J.~E., et~al. 2004, \apj, 611, 59
 
\bibitem[\protect\citeauthoryear{{Savaglio} et~al.}{{Savaglio}
  et~al.}{2005}]{sgl+05}
{Savaglio}, S., et~al. 2005, \apj, 635, 260
 
\bibitem[\protect\citeauthoryear{{Shapley} et~al.}{{Shapley}
  et~al.}{2005}]{scm+05}
{Shapley}, A.~E., {Coil}, A.~L., {Ma}, C.-P.,  \& {Bundy}, K. 2005, \apj, 635,
  1006
 
\bibitem[\protect\citeauthoryear{{Shapley} et~al.}{{Shapley}
  et~al.}{2004}]{sep+04}
{Shapley}, A.~E., {Erb}, D.~K., {Pettini}, M., {Steidel}, C.~C.,  \&
  {Adelberger}, K.~L. 2004, \apj, 612, 108
 
\bibitem[\protect\citeauthoryear{{Sirianni} et~al.}{{Sirianni}
  et~al.}{2005}]{sjb+05}
{Sirianni}, M., et~al. 2005, \pasp, 117, 1049
 
\bibitem[\protect\citeauthoryear{{Spergel} et~al.}{{Spergel}
  et~al.}{2006}]{sbd+06}
{Spergel}, D.~N., et~al. 2006, ArXiv Astrophysics e-prints
 
\bibitem[\protect\citeauthoryear{{Stanway} et~al.}{{Stanway}
  et~al.}{2004}]{sgb+04}
{Stanway}, E.~R., et~al. 2004, \apjl, 604, L13
 
\bibitem[\protect\citeauthoryear{{Starling} et~al.}{{Starling}
  et~al.}{2005}]{sve+05}
{Starling}, R.~L.~C., et~al. 2005, \aap, 442, L21
 
\bibitem[\protect\citeauthoryear{{Steidel} et~al.}{{Steidel}
  et~al.}{1999}]{sag+99}
{Steidel}, C.~C., {Adelberger}, K.~L., {Giavalisco}, M., {Dickinson}, M.,  \&
  {Pettini}, M. 1999, \apj, 519, 1
 
\bibitem[\protect\citeauthoryear{{Stern} et~al.}{{Stern} et~al.}{2005}]{sye+05}
{Stern}, D., {Yost}, S.~A., {Eckart}, M.~E., {Harrison}, F.~A., {Helfand},
  D.~J., {Djorgovski}, S.~G., {Malhotra}, S.,  \& {Rhoads}, J.~E. 2005, \apj,
  619, 12
 
\bibitem[\protect\citeauthoryear{{Stiavelli} et~al.}{{Stiavelli}
  et~al.}{2005}]{sdp+05}
{Stiavelli}, M., et~al. 2005, \apjl, 622, L1
 
\bibitem[\protect\citeauthoryear{{Tagliaferri} et~al.}{{Tagliaferri}
  et~al.}{2005}]{tac+05}
{Tagliaferri}, G., et~al. 2005, \aap, 443, L1
 
\bibitem[\protect\citeauthoryear{{Taniguchi} et~al.}{{Taniguchi}
  et~al.}{2005}]{tan+05}
{Taniguchi}, Y., et~al. 2005, \pasj, 57, 165
 
\bibitem[\protect\citeauthoryear{{Totani} et~al.}{{Totani}
  et~al.}{2005}]{tkk+05}
{Totani}, T., {Kawai}, N., {Kosugi}, G., {Aoki}, K., {Yamada}, T., {Iye}, M.,
  {Ohta}, K.,  \& {Hattori}, T. 2005, ArXiv Astrophysics e-prints
                                                                                                              
\bibitem[\protect\citeauthoryear{{Tremonti} et~al.}{{Tremonti}
  et~al.}{2004}]{thk+04}
{Tremonti}, C.~A., et~al. 2004, \apj, 613, 898
 
\bibitem[\protect\citeauthoryear{{Wainwright}, {Berger}, \&
  {Penprase}}{{Wainwright} et~al.}{2005}]{wbp05}
{Wainwright}, C., {Berger}, E.,  \& {Penprase}, B.~E. 2005, ArXiv Astrophysics
  e-prints
 
\bibitem[\protect\citeauthoryear{{Westra} et~al.}{{Westra}
  et~al.}{2005}]{wjl+05}
{Westra}, E., et~al. 2005, \aap, 430, L21
 
\bibitem[\protect\citeauthoryear{{Yan} \& {Windhorst}}{{Yan} \&
  {Windhorst}}{2004}]{yw04}
{Yan}, H.,  \& {Windhorst}, R.~A. 2004, \apjl, 600, L1
 
\end{thebibliography}
\end{document}